# Enhancement of optical response in nanowires by negative-tone PMMA lithography

Ilya Charaev, Andrew Dane, Akshay Agarwal, Karl K. Berggren

*Abstract*— **The method of negative-tone-PMMA electron-beam lithography is investigated to improve the performance of nanowire-based superconducting detectors. Using this approach, the superconducting nanowire single-photon detectors (SNSPDs) have been fabricated from thick 5-nm NbN film sputtered at the room temperature. To investigate the impact of this process, SNSPDs were prepared by positive-tone and negative-tone-PMMA lithography, and their electrical and photodetection characteristics at 4.2 K were compared. The SNSPDs made by negative-tone-PMMA lithography show higher critical-current density and higher photon count rate at various wavelengths. Our results suggest a higher negative-tone-PMMA technology may be preferable to the standard positive-tone-PMMA lithography for this application.**

## I. Introduction

Since the original publication of Haller *et al*. in 1968 [1], polymethyl methacrylate (PMMA) has been used as one of the most common positive-tone resists for high-resolution electron-beam lithography. The positive nature of this resist occurs due to polymer chain scissions which are initiated by the electron beam. PMMA resist can be used to reproducibly obtain patterns with nm sizes. The resolution can be further improved by adjusting different steps of the lithographic process such as the writing strategy or the development process. In recent decades, HSQ (hydrogen silsesquioxane) negative-tone resist has become a serious candidate for a high-resolution electron-beam resist because of its small line edge roughness, high etching resistance and small molecular size [2]. However, after exposure and development, the etch resistance of crosslinked HSQ increases [3], and its subsequent removal requires a hydrofluoric acid dip or a $CF_4$ reactive ion etch (RIE); both of which can damage an underlying metallic thin film. HSQ is also notoriously unstable in the presence of water vapor, making its processing problematic in many lab environments. As an alternative, PMMA also be used as a negative resist when the polymerization becomes more dominant at high exposure doses [4]. Although the pitch of the dense features is smaller than what can be obtained with PMMA positive tone, the stability of the structures is a serious problem when the aspect ratio is high [5].

Ilya Charaev, Andrew Dane, Akshay Agarwal, Karl K. Berggren are with the Department of Electrical Engineering and Computer Science, Massachusetts Institute of Technology, Cambridge, MA 02139, USA (e-mail: charaev@mit.edu).

The process of electron-beam lithography is a crucial for fabrication of superconducting nanowires. Line edge roughness becomes a serious issue when the pattern size shrinks [6]. For nanolithography, the line edge roughness should be as small as possible in order to avoid pattern distortion or deterioration of the resolution.

Experiment suggests [7] that the critical current of the fabricated devices is affected by nanowire edge defects (so-called lateral proximity effect) and by internal structural defects that weaken the superconducting order parameter. To minimize the suppression of the critical current in nanowires, the size and frequency of defects should be reduced. Among the high-resolution electron beam resists, only a few of them allow us to fabricate structures with sub-10-nm half-pitch resolution [8]. Although the highest resolution of positive polymethyl methacrylate (PMMA) is limited at 10-15 nm half-pitch, the PMMA resist operating in negative tone may permit a smaller half-pitch [9].

In past work, a series of single-nanowires were prepared by using positive-tone and negative-tone PMMA lithography. These nanowires were then compared in terms of critical temperature $T_C$, critical current density $j_C$ and retrapping current $I_r$ for widths from 50 nm up to 20 μm [10]. All observed metrics were found to be higher for sub-micrometer nanowires made by negative-tone PMMA lithography. The reproducible procedure of nanowire patterning proposed in [10] thus enabled improvement of the superconducting characteristics of NbN nanowires. However, single-photon detection with the fabricated nanowires was never demonstrated and thus the impact on device performance was not confirmed.

In this paper, we show that the usage of negative-PMMA lithography for fabrication of SNSPDs leads not only to enhanced critical current in nanowires, but also that it directly increases the photon count rate. We systematically studied the performance of detectors made by positive- and negative-PMMA lithography. Obtained results suggest that negative-tone PMMA lithography is a promising technique for making high-performance devices.

## II. Technology

The devices were fabricated according to a relatively standard materials growth process, followed by characterization x-ray reflection and diffraction, transmission-electron microscopy, and $T_C$ measurement prior to patterning and etch.

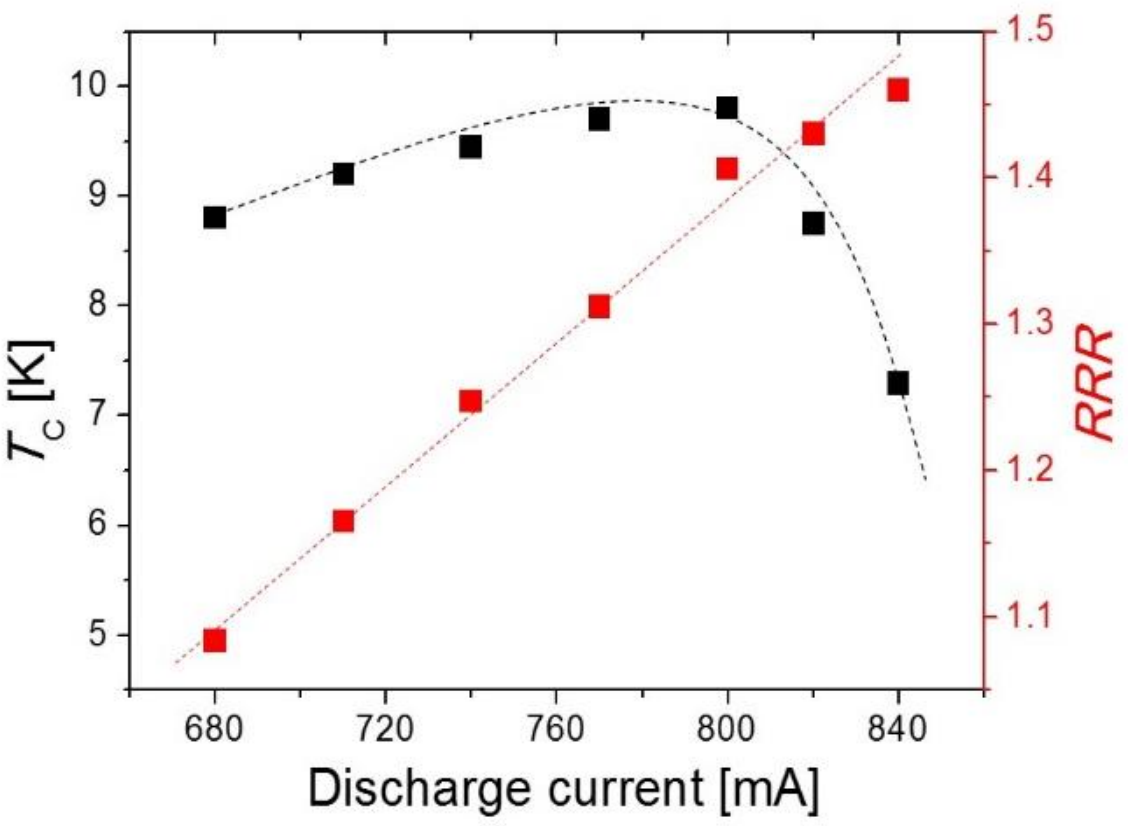

Fig. 1. The critical temperature and residual resistance ratio of thick 5-nm NbN film as a function of discharge current.

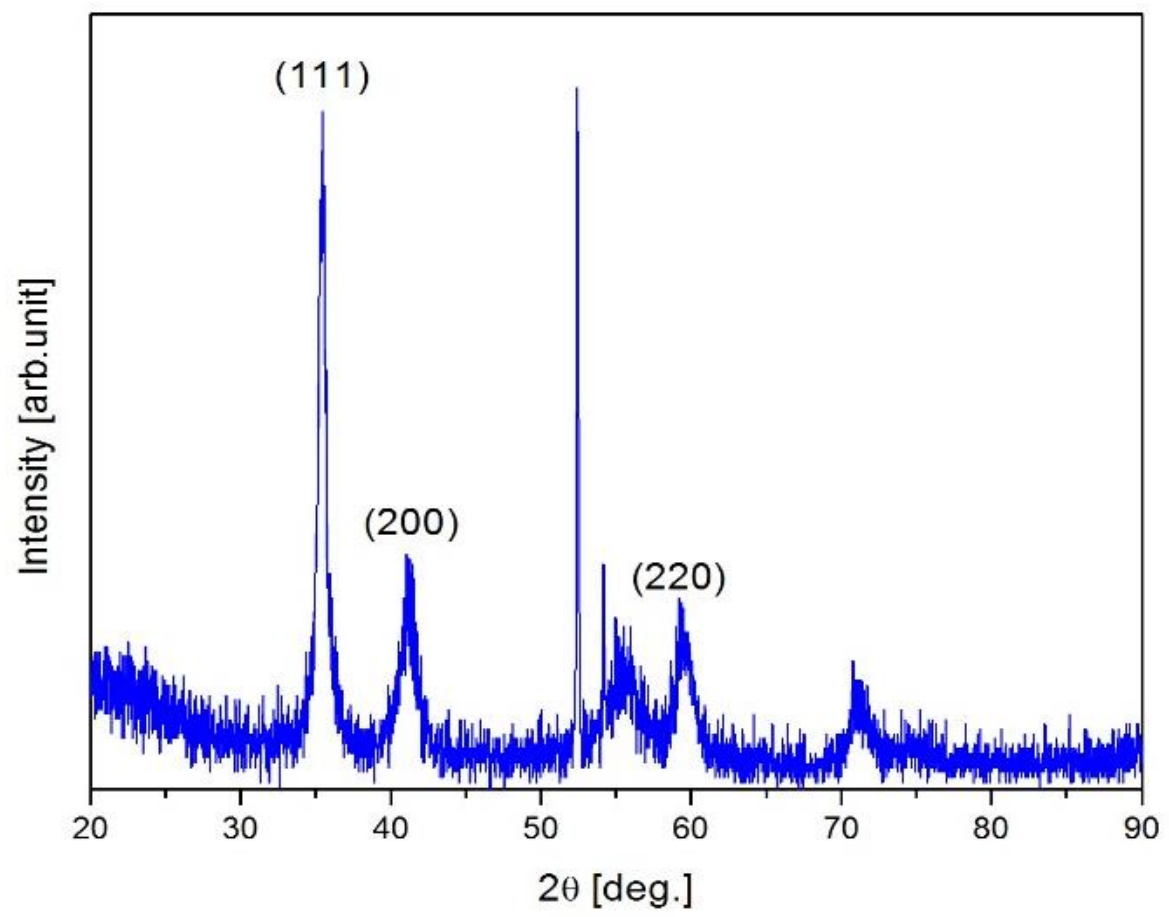

Fig. 3. X-ray diffraction from a ~50 nm thick NbN film deposited at the same condition as a 5 nm film.

### A. Thin-film deposition

The DC reactive magnetron sputtering process of NbN film was optimized using commercial sputtering system (AJA International Inc Orion series). The NbN films were deposited from a 2-inch-diameter pure Nb target (99.95%) in an argon (Ar) and nitrogen ($N_2$) atmosphere on magnesium oxide substrates (MgO) in current bias mode. The surface of the Nb target was pre-cleaned in a pure Ar atmosphere at 2.5 mTorr. Prior to the deposition, the plasma was stabilized in mixed Ar and $N_2$ atmosphere. An RF power of 14 W was applied to the 100 mm diameter sample holder from an external RF source and matching network in order to sputter-clean samples prior to deposition. The deposition was changed from processes used previously [11]. The range of deposition was shifted to a higher discharge current with increase the argon flow (31.5 sccm). Also, a 50 W RF bias wasn't added to the substrate holder during sputtering.

Thicknesses were measured using x-ray reflectivity (XRR) to determine the deposition rate for each of the conditions studied. At each chamber condition, the sputtering time was subsequently adjusted to yield 5-nm thick films. Electrical characterization of the thin films included four-point probe measurements of film $T_C$ and sheet resistance. The measurements were performed in the temperature range from 5.5 up to 300 K. The critical temperature was defined at the point where $R = R_{20K}/2$.

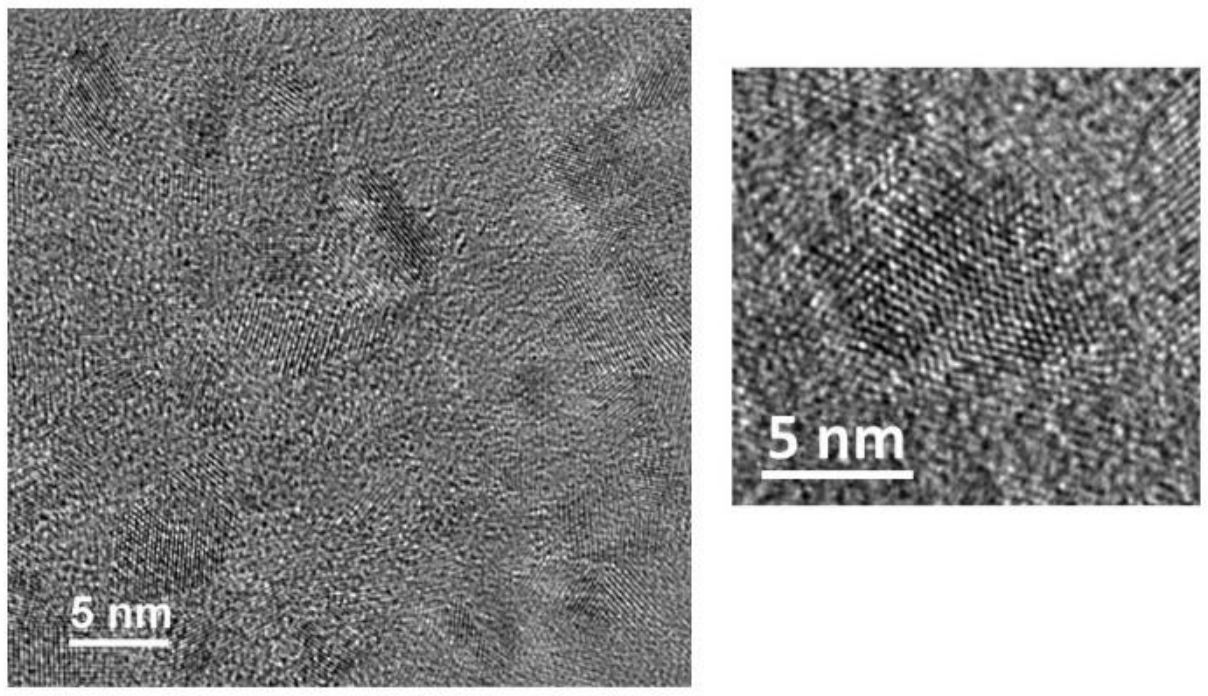

Fig. 2. Image of film surface of thick 5-nm NbN film obtained by TEM (left image). Right graph represents smaller region of NbN film.

The residual-resistance ratio *RRR* was calculated as ratio $R_{300K}/R_{20K}$. The resistivity $\rho_0$ was evaluated using the measured thickness $d$ and the sheet resistance of the film at 25 K as: $\rho_0 = R_{20K} \times d$.

Figure 1 represents the critical temperature (black points) and residual-resistance ratio (red points) as function of discharge current for thick 5-nm NbN. The maximum $T_C$ value (9.8 K) was obtained for films deposited at 800 mA. Both increase and decrease of the sputter current from this value results in a reduced critical temperature and this reduction aws stronger towards the Nb phase. This is further confirmed by the increase of the residual resistivity ratio. The deposition rate was 0.14 nm/s at 800 mA.

The crystal structure of film sputtered at 800 mA was characterized by X-ray diffraction (XRD) analysis (Rigaku XRD SmartLab) with attenuator correction (high resolution monochromator PB-Ge(220)×2) on thick 50 nm NbN film sputtered at the same conditions. According to XRD data displayed on the graph in Fig. 3, the film was polycrystalline in nature and showed a preferred orientation along the (111) plane; other weak peaks are (200) and (220) [12]. These peaks appear to be consistent with cubic $NbN_{0.88}$.

Figure 2 shows TEM images of a 5 nm thick NbN film obtained by JEOL 2010F which confirms the polycrystalline structure of our films. TEM imaging was performed on a specially prepared NbN film which was sputtered onto a commercially available, 20 nm thick $Si_3N_4$ window. The TEM window was oxygen plasma cleaned ex-situ before being installed in the sputtering chamber. In-situ sputter cleaning was used for all TEM samples with the same settings as used for the other samples.

### B. Devices fabrication

The SNSPDs were fabricated using positive-tone and negative-tone PMMA technology described in detail in [10]. We used a Raith 150 EBL tool at 10 kV acceleration voltage, 20 μm aperture, resulting in a beam current of ~120 pA. The film was patterned into detectors *via* electron-beam lithography over the PMMA resist and subsequently reactive ion etching in $CF_4$ at

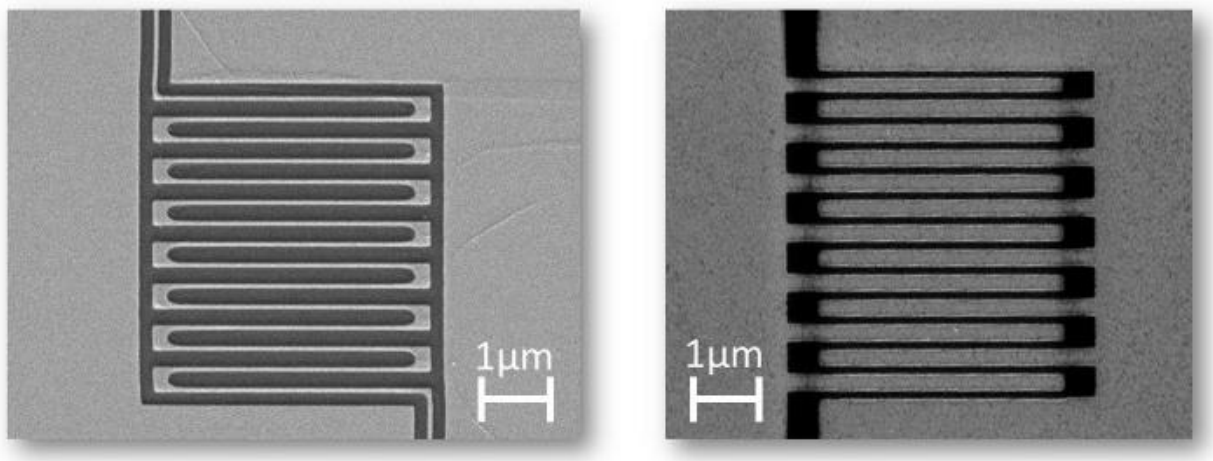

Fig. 4. SEM images meander-type SNPSD made by positive- (left) and negative-PMMA (right) lithography.

50 W. The nanowires were exposed by using a dose of 100 µC/cm$^2$ and 10000 µC/cm$^2$ for positive and negative-tone lithography respectively. The negative pattern was developed in acetone for 1.5 min and then rinsed in 2-propanol while the standard developer for 85 s (30% MIBK in 2-propanol at 22°C) was used for developing of positive pattern.

To perform statistical analysis of SNSPDs made by both techniques, 25 samples of each technique were made on the same 10×10 mm$^2$ chip. The example of detector layout is presented on Fig. 4 for positive- (left) and negative-PMMA (right) lithography. The detector area is 4×4 µm$^2$ for all specimen. The width of nanowires $W$ is 100 nm with 300 nm gap to minimize current-crowding [13].

## III. Measurements

The resulting devices were characterized by DC electrical measurements and by optical testing with light at a range of wavelengths.

### A. *Experimental setup*

The measurements were performed in a standard 100-liter $He^4$ transport dewar and operated at the temperature 4.2 K. The chip was glued on the holder. The detectors were bonded with aluminum wires to contact pads on the holder. The high-frequency and DC signals were led out of the dipstick by stainless-steel rigid coaxial cables. The room temperature bias tee decoupled the high frequency path from the DC bias path. The signal was amplified at room temperature by a 60-dB amplifier and then sent to a pulse counter with 500 MHz internal bandwidth. The samples were biased by a low-noise voltage source. The detectors were optically characterized by laser light at 780 nm and 1550 nm wavelengths. The power was attenuated to ensure the single-photon counting regime of detection.

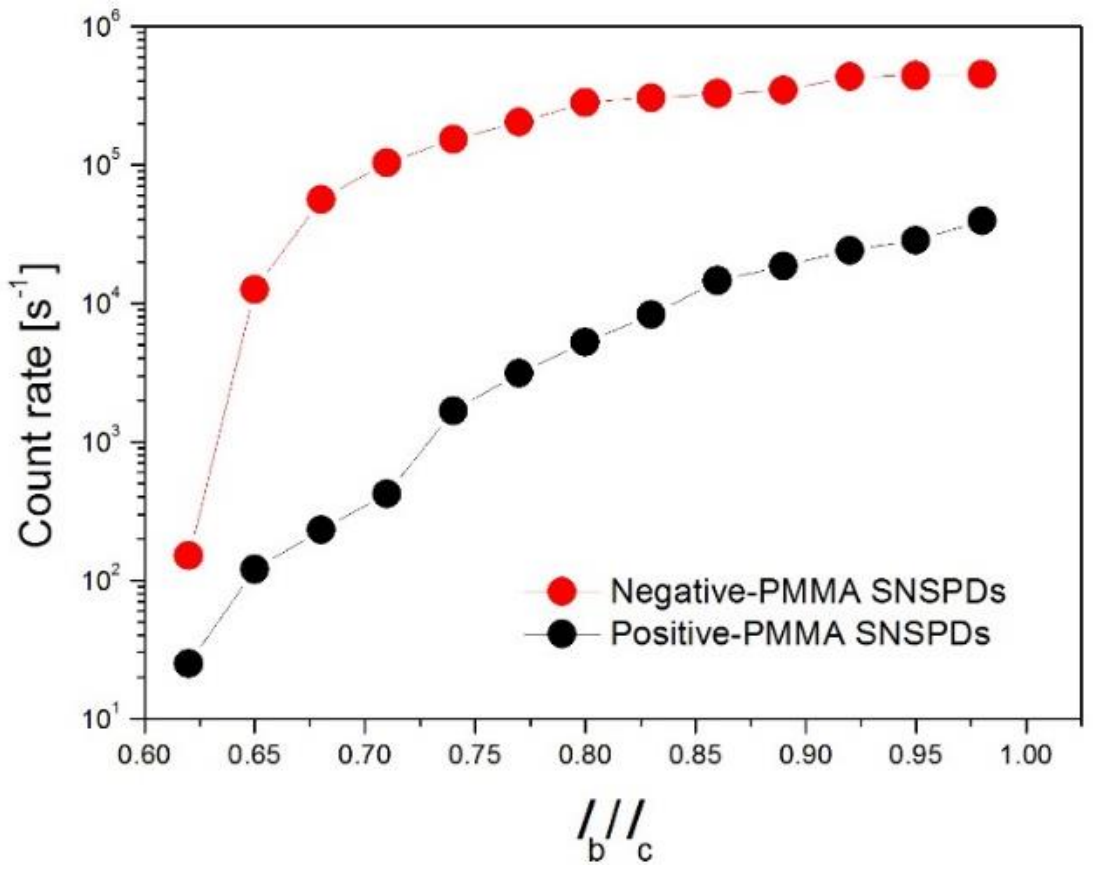

Fig. 6. The count rate as a function of the relative bias current at 780 nm wavelength for positive- (black) and negative-PMMA (red) SNSPD.

### B. *Results*

The current-voltage characteristics of all SNSPDs were measured at $T$ = 4.2K. The critical current $I_C$ is associated with the step-like switching of the nanowire from the superconducting to the normal state when the current increases from zero. The nominal density of the critical current at $T$ = 4.2 K was calculated as $j_C(4.2\text{ K}) = I_C(4.2\text{ K})/Wd$, where $W$ is a width measured by SEM.

The Fig. 5 represents the statistical distribution of the critical current density of SNSPDs prepared by both techniques. The calculated values of $j_C$ for positive-tone PMMA SNSPDs were found between 7.2 and 8.3 MA/cm$^2$. The negative-tone PMMA SNSPDs demonstrated a higher critical current density, in the range of 9.2-10 MA/cm$^2$. The average difference in $j_C$ was ~ 30% between positive- and negative-PMMA SNSPDs.

Dark- and photon-count rate measurements were performed for both types of SNSPD in a single cooling cycle. First, the count rate as a function of the relative current was measured under irradiation at 780 nm. Fig. 6 shows an example of the optical response obtained for positive- and negative-PMMA SNSPDs. Although the detection onset appears at the same relative bias current ($I_b/I_C$=0.63) for both types of SNSPDs, the count rate of negative-tone PMMA was significantly higher. The counts of negative-tone SNSPDs were saturated at the bias current close to the critical value, while no saturation was observed for all SNSPDs made by positive-tone PMMA lithography.

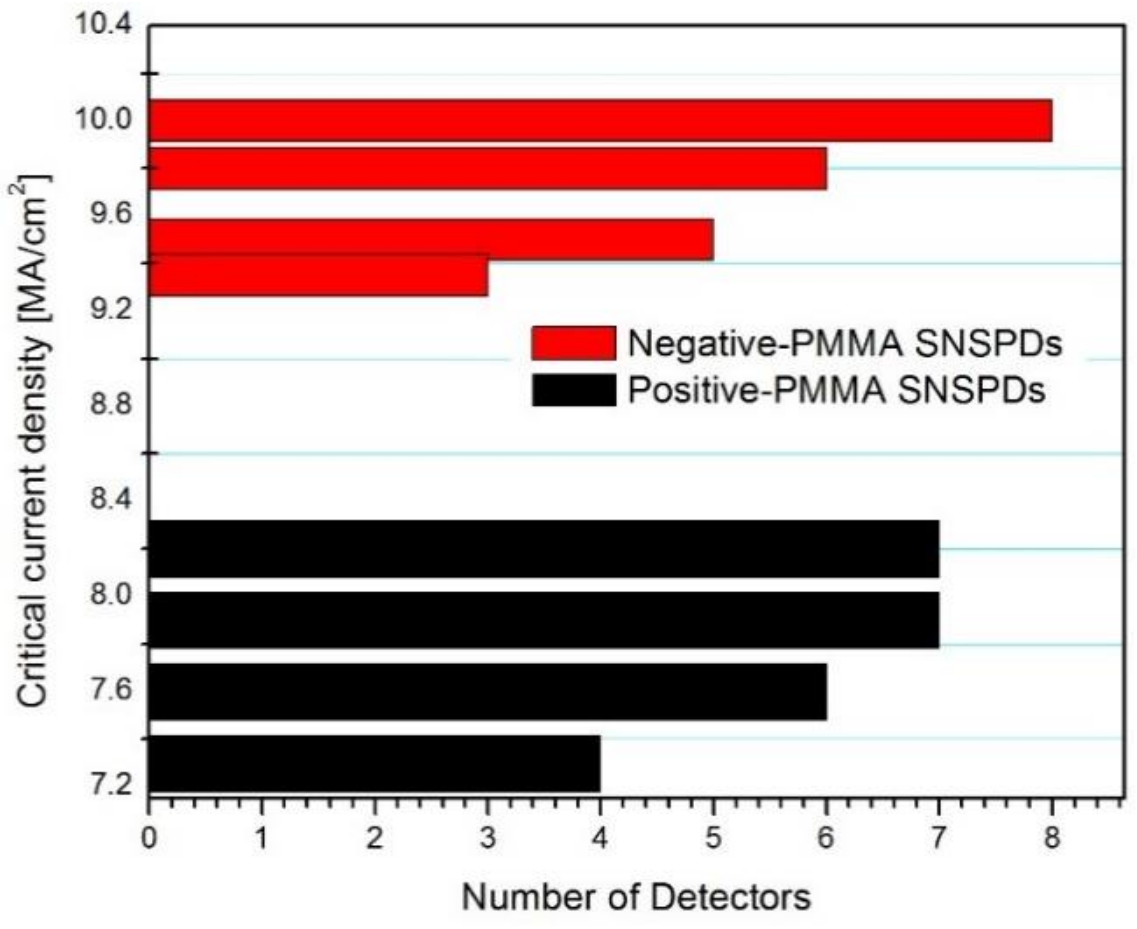

Fig. 5. Statistical distribution of critical current density of SNSPDs made by positive- (black) and negative-PMMA (red) lithography.

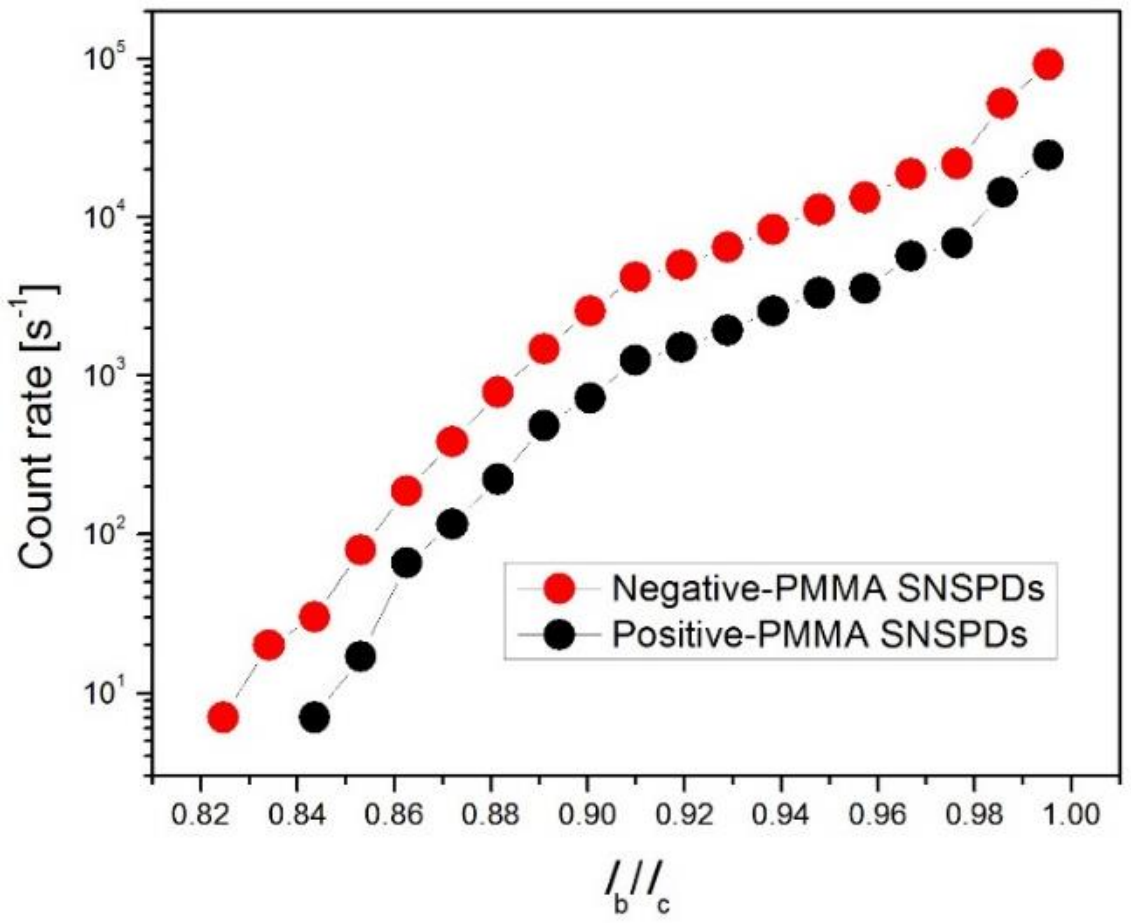

Fig. 7. The count rate as a function of the relative bias current at 1550 nm wavelength for positive- (black) and negative-PMMA (red) SNSPD.

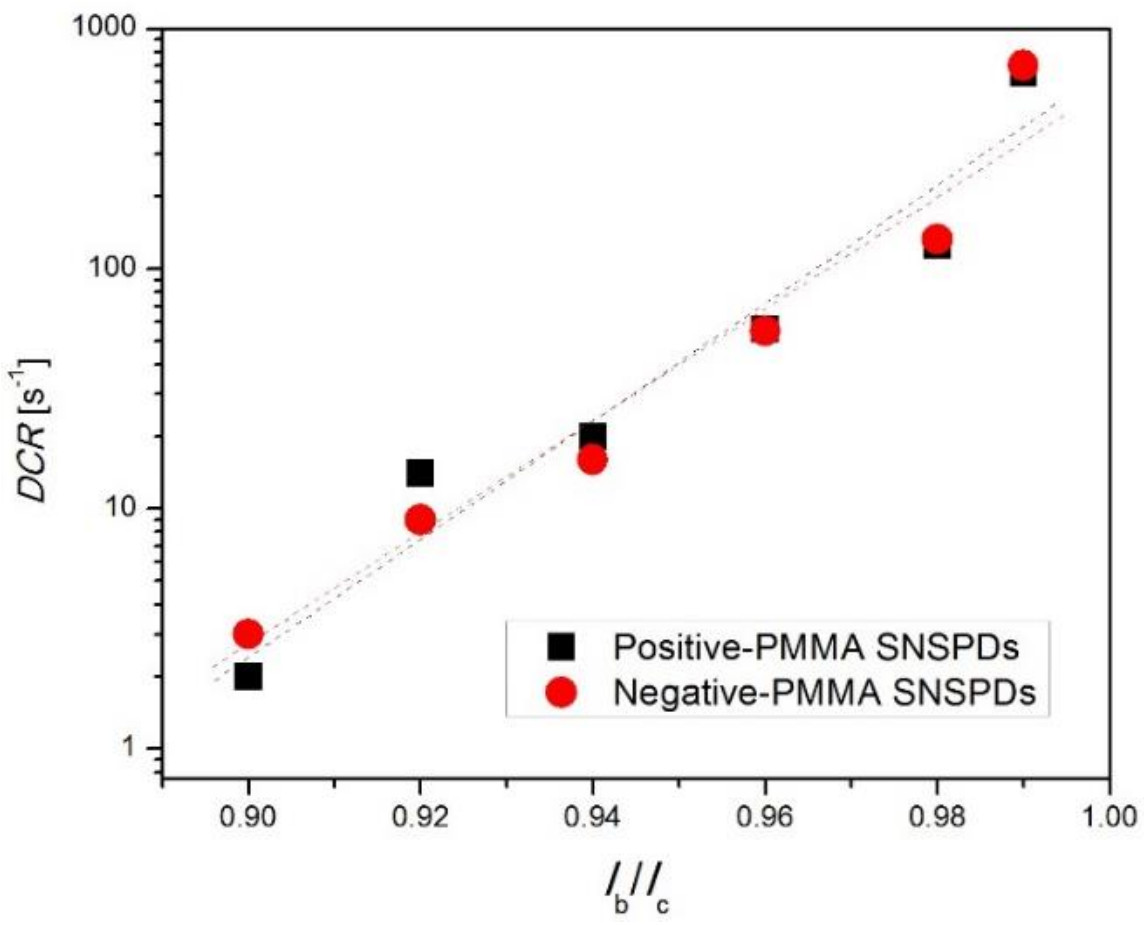

Fig. 8. The dark count rate as a function of the relative bias current plotted for positive- (black) and negative-PMMA (red) SNSPD.

Identical measurements were then done at 1550 nm wavelength. Both types of detectors show similar count-rate dependence (Fig. 7). Despite the absence of saturation, the negative-PMMA SNSPDs demonstrated higher count rate in comparison with the positive-tone PMMA detectors. Starting at $0.82I_C$, the count rate grew with increase in the current for both types SNSPDs.

A similar difference in count rate at 780 nm and 1550 nm wavelengths was observed for all measured SNSPDs (5 detectors of each type). Moreover, the saturation of the internal efficiency was found only for negative-tone PMMA SNSPDs.

The dark-count rate *DCR* was determined by blocking the optical path at the warm end of fiber. Detectors had similar dark count rates independent of the type of SNSPD. An example of *DCR* as function of relative bias current is presented on Fig. 8. The dark counts were observed at currents of $0.9I_C$ and higher. The maximum *DCR* was measured as $<10^3$ cps at current closed to the critical value.

### C. Discussion

An enhancement of critical current density and photon count rate in negative-tone PMMA SNSPDs was observed. The critical current density of negative-tone PMMA SNSPDs was 30% higher than the $j_C$ of positive-tone PMMA detectors with identical geometry. Because the photon count rate of SNSPDs grows with increase of absolute bias current, we presume the higher photon count rate observed for the negative-tone process was a simple consequence of the increased bias current.

The observation of higher count rates and saturation at 780 nm measured at 4.2 K for the negative-tone process suggests that the critical current of negative-tone devices approaches the depairing current in nanowires more closely than the positive-tone devices. Different models of SNSPD response predict dependence of the red boundary on ratio of bias to de-paring current [14]. The higher critical current density of the negative-tone devices is expected to also result in decrease of the minimal detectable energy of photon.

The rate of dark counts was found to be similar for both types of SNSPDs. This may be surprising, given that the negative-tone devices had much higher critical-current density. However, it has been argued that dark counts in these devices originate at bends and turns [15]. The layouts of meanders for positive- and negative-PMMA detectors contained equal numbers of bends, which may explain the similar dark-count-rates observed in the two experiments.

Although the layout was optimized to reduce the current crowning in bends, the meander was not bend-free. To push the critical current density closer to its depairing value, the detector shape should be optimally curved [13].

## IV. Conclusion

We performed comparisons of SNSPDs made by positive- and negative-tone PMMA technology. Using negative-tone-PMMA lithography for fabrication of SNSPDs, the performance of detectors was enhanced in terms of its critical current and count rate at different wavelengths. The demonstrated method permits reproducible fabrication of nanowires with enhanced critical current density.

## Acknowledgment

The authors would like to thank Jim Daley and Mark Mondol of the MIT Nanostructures lab for the technical support related to electron beam fabrication. The authors thank Dr. C.-J. Chung for helpful discussion. They also like to thank Dr. Charles Settens from the Center for Materials Science and Engineering X-ray Facility for his assistance and advice on all matters related to x-ray measurements. Support for this work was provided in part by the DARPA Defense Sciences Office, through the DETECT program, MIT-Skoltech Program and program “Process & Materials Innovation for Superconducting Nanowire Single-Photon Detectors”.

## References

[1] I. Haller, M. Hatzakis and R. Shrinivasan, "High-resolution Positive Resists for Electron-beam Exposure," *BM J. Res.*, vol. 12, no. 3, pp. 251 - 256, May 1968.

[2] H. Namatsu, Y. Takahashi, K. Yamazaki, T. Yamaguchi, M. Nagase, and K. Kurihara, "Three-dimensional siloxane resist for the formation of nanopatterns with minimum linewidth fluctuations," *J. Vac. Sci. Technol. B*, vol. 16, no. 69, June 1998.

[3] M. Rommel and J. Wei, "Hydrogen silsesquioxane bilayer resists—Combining high resolution electron beam lithography and gentle resist removal," *J. Vac. Sci. Technol. B*, vol. 06F102, Oct. 2013.

[4] I. Zailer, J. Frost, V. Chabasseur-Molyneux, C. Ford and M. Pepper, "Crosslinked PMMA as a high-resolution negative resist for electron beam lithography and applications for physics of low-dimensional structures," *Semicond. Sci. Technol.*, vol. 11, no. 1235, Apr. 1996.

[5] A. Hoole, M. Welland and A. Broers, "Negative PMMA as a high-resolution resist—the limits and possibilities," *Semicond. Sci. Technol.*, vol. 12, no. 1166, June 1997.

[6] H. Namatsu, Y. Takahashi, K. Yamazaki, T. Yamaguchi, M. Nagase, and K. Kurihara , "Three-dimensional siloxane resist for the formation of nanopatterns with minimum linewidth fluctuations," *J. Vac. Sci. Technol. B*, vol. 16, no. 69, Nov. 1997.

[7] I. Charaev,, T. Silbernagel, B. Bachowsky, A. Kuzmin, S. Doerner, K. Ilin, A. Semenov, D. Roditchev, D. Yu. Vodolazov, and M. Siegel, "Proximity effect model of ultranarrow NbN strips," *Phys. Rev. B*, vol. 96, no. 184517, Nov. 2017.

[8] A. Grigorescu and C. Hagen, "Resists for sub-20-nm electron beam lithography with a focus on HSQ: state of the art," *Nanotechnology*, vol. 20, no. 292001, Feb. 2009.

[9] H. Duan, D. Winston, J. K. Yang, B. Cord, V. Manfrinato, and K. Berggren, "Sub-10-nm half-pitch electron-beam lithography by using poly(methyl methacrylate) as a negative resist," *J. Vac. Sci. Technol. B*, vol. 28, no. C6C58, Sep. 2010.

[10] I. Charaev, T. Silbernagel, B. Bachowsky, A. Kuzmin, S. Doerner, K. Ilin, M. Siegel, A. Semenov, D. Roditchev, and D. Yu. Vodolazov, "Enhancement of superconductivity in NbN nanowires by negative electron-beam lithography with positive resist," *J. Appl. Phys.*, vol. 122, no. 083901, Aug. 2017.

[11] A. Dane, A. McCaughan, D. Zhu, Q. Zhao, C.-S. Kim, N. Calandri, A. Agarwal, F. Bellei, and K. Berggren, "Bias sputtered NbN and superconducting nanowire devices," *Appl. Phys. Lett.*, vol. 111, no. 122601, Sep. 2017.

[12] A. Miura, T. Takei, N. Kumada, S. Wada, E. Magome, C. Moriyoshi, and Y. Kuroiwa, "Bonding Preference of Carbon, Nitrogen, and Oxygen in Niobium-Based Rock-Salt Structures," *Inorg. Chem.*, vol. 52, no. 17, p. 9699–9701, Aug. 2013.

[13] John R. Clem and Karl K. Berggren, "Geometry-dependent critical currents in superconducting nanocircuits," *Phys. Rev. B*, vol. 84, no. 174510, Nov. 2011.

[14] A. Engel, J. J. Renema, K. Ilin, A. Semenov, "Detection mechanism of superconducting nanowire single-photon detectors," *Supercond. Sci. Technol.*, vol. 114003, no. 28, Sep. 2015.

[15] A. Semenov, I. Charaev, R. Lusche, K. Ilin, M. Siegel, H.-W. Hübers, N. Bralović, K. Dopf, and D. Yu. Vodolazov, "Asymmetry in the effect of magnetic field on photon detection and dark counts in bended nanostrips," *Phys. Rev. B* , vol. 92, no. 174518, Nov. 2015.